\documentclass[preprint2]{aastex62}	 
\usepackage[T1]{fontenc}
\usepackage{xcolor}
\usepackage[normalem]{ulem}
\usepackage[caption=false]{subfig}

\shortauthors{Brittain et al.}
\shorttitle{HD~179218}

\begin{document}
\title{Spectroastrometric Study of Ro-vibrational CO Emission from the Herbig Ae star HD~179218 with iSHELL on the NASA Infrared Telescope Facility}

\author{Sean D. Brittain}
\affil{Department of Physics \& Astronomy, 118 Kinard Laboratory, Clemson University, Clemson, SC 29634-0978, USA }
\email{sbritt@clemson.edu}

\author{John S. Carr}
\affil{Naval Research Laboratory, Code 7211, Washington, DC 20375, USA}

\author{Joan R. Najita}
\affiliation{National Optical Astronomy Observatory, 950 N. Cherry Ave., Tucson, AZ 85719, USA}

\begin{abstract}
We present analysis of commissioning $M-$band data acquired with the infrared echelle spectrograph (iSHELL) on {\it NASA's Infrared Telescope Facility}. In this paper we describe the delivered performance of the instrument for these $M-$band observations and the data reduction process. The feasibility of using iSHELL for spectro-astrometry is tested on the Herbig Ae/Be star HD~179218 and we show that sub-milliarcsecond fidelity is achievable. 
\end{abstract}

\section{Introduction}
High-resolution spectroscopic observations of CO in disks around young stars have been used to elucidate the evolutionary status of warm gas in disks around young stars \citep{najita2000, najita2003, najita2007, carr2001, carr2007, brittain2003, brittain2007, blake2004, salyk2009, salyk2011, bast2011, herczeg2011, vdp2015, banzatti2015, hein2016}. The application of spectro-astrometry (SA) to this observational technique provides additional spatial information about the disk on milli-arcsecond scales \citep{pontoppidan2008}. It has enabled the identification of non-axisymmetric structures in disks that may point to the presence of disk winds \citep{ pontoppidan2011} and forming gas giant planets \citep{brittain2013}. 

SA is the measurement of the 
centroid of the point spread function (PSF) of a 
spectrum as a function of wavelength or velocity. SA has been 
used to study phenomena such as binaries 
\citep{bailey1998, baines2004, porter2004}, outflows 
\citep{whelan2008, davies2010}, and disks \citep[e.g.][]
{acke2006, brittain2009, pontoppidan2011}.  
While the spatial resolution of a PSF is limited to 
$\sim$1.2D/$\lambda$, the relative centroid of the PSF 
can be measured to a small fraction of that. For a well 
sampled Gaussian PSF dominated by photon 
noise, the center of the Gaussian can be measured to 
an accuracy of $\sim$0.4$\times$FWHM/SNR \citep{brannigan2006} 
where FWHM is the full-width of the PSF at half of its maximum, and 
SNR is the per pixel signal to noise ratio of the continuum. 
For FWHM=0$^{\arcsec}$.7 and SNR = 100, we can measure the centroid of each velocity 
channel of the line to an accuracy of 3 milliarcseconds 
(~1AU at the distance of HD~179218). By combining 
multiple lines, this fidelity can be pushed down further. 

An important consideration with SA observations is 
correction for artifacts \citep[][and references therein]{brittain2015}. 
By observing each desired position angle with an 
anti-parallel observation, most artifacts can be 
removed in the data processing. However, grating 
and slit stability are important to ensure that 
artifacts are reproducible. Here we present analysis 
of iSHELL commissioning data to test the feasibility 
of using this instrument for SA measurements. 

iSHELL is a cross-dispersed echelle spectrograph sensitive from 
1-5$\mu$m on the NASA Infrared Telescope Facility (IRTF). The 
IRTF is a 3m 
telescope optimized for observations in the infrared. With a 
0$^{\arcsec}$.375 slit, the nominal spectral resolution of iSHELL is $\lambda/\Delta\lambda=75,000$.  
iSHELL has a 2k H2RG detector that provides coverage of nearly the entire 
$M-$band in a single setting \citep{rayner2016}. Here we present a study
of ro-vibrational CO emission from the Herbig Ae/Be star HD~179218. 

HD~179218 is a Meeus Group 1 \citep{meijer2008} Herbig~Ae/Be 
star with a transition disk 
\citep[e.g.,][]{menu2015}. The distance to HD~179218 is 
293.1$\pm$30.5~pc 
\citep{GC2, GC1}.This is 20\% larger than the distance 
inferred from Hipparchus data (253$\pm$38~pc; 
\citealt{vanleeuwen2007}) and 50\% larger than the distance 
inferred from photospheric modeling of the star 
(201$^{-22}_{+25}$~pc; \citealt{montesinos2009}). Where 
appropriate, we rescale system parameters taken from the 
literature using the distance reported from Gaia. 

Fitting the Balmer lines from the stellar photosphere 
indicates that T$\rm_{eff}$=9640$\pm$250~K and log(g)=3.9$\pm$0.2 
\citep{folsom2012}. The stellar luminosity is 133$\pm$28 L$_{\sun}$. 
Using the Siess tracks, this stellar luminosity and 
effective temperature is consistent with a stellar 
mass of 3.0$\pm$0.1 M$_{\sun}$ and stellar radius of 
3.3$\pm$0.3 R$_{\sun}$ \citep{siess2000}. 

Adopting the 1.3mm flux reported by \citep{aa2009}, assuming a disk 
temperature of 50~K and gas-to-dust mass ratio of 100, the disk mass is 
0.07M$_{\sun}$. The accretion rate is 1.7$\times$10$^{-8}$M$\rm_{\odot}$
yr$^{-1}$ \citep{mendigutia2017}, and the accreting material is depleted 
in refractory material indicating that the ratio of gas to dust of the 
accreting material is $\sim$400 \citet{kama2015}. This system is 
consistent with a transition disk opened by a massive companion. 

 The inclination of the disk is uncertain. Modeling of the rotational CO
 line profiles suggest the inclination is 40$\pm$10$^{\degr}$ 
 \citep{dent2005}. VLTI/PIONEER data is consistent with a disk 
 inclination of 49$^{\degr}$ \citep{lazareff2017}. Fitting the $N-$band 
 visibilities measured with VLTI/MIDI indicate a disk inclination of 
 57$\pm$2$^{\degr}$ and position angle of 23$\pm$3$^{\degr}$.
 
 In this paper we describe the observations and data reduction (section 2). Next we characterized the reduced spectra and spectroastrometric signal (section 3). We compare the data acquired with iSHELL to CRIRES obervations of the same source (section 4), and conclude with an analysis of the excitation of the gas (section 5) and suggest future work. 

\section{Observations}
The data were acquired October 11, 2016 as part of the science 
commissioning program for iSHELL. The observation log is summarized 
in Table 1. A total of 30 minutes of data were taken using the M2 
grating setting and a slit width of  0$^{\arcsec}$.375. The observations 
were taken in an ABBA nod pattern in order to cancel the sky background 
to first order. At each position, the integration time was 15 seconds 
and 2 coadds. The observations were divided into two sets. For the 
first 15 minutes of integration, the position angle was fixed at 
23$^{\degr}$ east of north. The slit was then rotated to 203$^{\degr}$ 
east of north for the second half of the observations. A telluric 
standard, HD~177724, was observed at a similar airmass. Flats were 
taken immediately following the observation before the grating was moved. 

The two sets of data from HD179218 and the telluric standard were 
reduced individually using interactive software written in IDL. The 
median of the ``A'' files and the median of the ``B'' files were used 
to flag cosmic ray hits. The non-flagged pixels were then averaged. 

The M2 setting includes 16 orders covering 1910.6 - 2217.5 cm$^{-1}$. 
The coverage of the orders ranges from $\sim$16~cm$^{-1}$ at the red 
end to 18.5~cm$^{-1}$ at the blue end. The gap between orders ranges 
from 3~cm$^{-1}$ at the red end to 1~cm$^{-1}$ at the blue end. Thus 
the total coverage of the M-band is $\sim$90\%. 

Custom software was written in IDL for reducing this data following
the process described in \citep{brittain2004}. Our approach differs
from the iSHELL tool in that each order is reduced separately and the 
rectified 2-D spectra are saved for SA measurements. To reduce the
data, the user selects several points along the A and B beam, 
and a polynomial is fit to these points. The mid-point of the PSF
is defined for each column of the order and 
used to create a box to extract the order to be reduced. This region 
is divided by the normalized flat and the A and B frames are 
subtracted to produced a flat-fielded sky subtracted image. 

A refined trace along the spectra in the order is found by fitting 
a Moffat function to the PSF in each column using MPFIT \citep{mark2009}. 
The bad pixel mask provided by the observatory is used to exclude bad 
pixels from the fit. Columns where the fit diverges by more than 3 
pixels from the centroid defined by the user are excluded (this is 
usually because there is little continuum in a given column). Deviant 
pixels are flagged and replaced by the value of the fit. A third-degree 
polynomial is fit to the center of the PSF in each column and the rows 
are shifted so that the A and B beams are rectified on the array. 

The same shift is applied to the non-sky subtracted A and B frames. 
The top half of the A-frame and bottom half of the B-frame (where 
there is no stellar spectrum) are combined to create a sky emission 
frame. Each row of this frame is fit with radiance model generated 
by the Sky Synthesis Program \citep[SSP;][]{kunde1974} which accesses the 
HITRAN2000 molecular line database \citep{rothman2003}. The first 
row is fit interactively, then the wavelength solution is refined 
by optimizing the fit of the model to each row of the sky spectrum. 
The rows are then interpolated to the solution of the middle row 
so that the spectra are rectified in the spectral dimension as well. 

The A and B beam are extracted using rectangular extraction of the 
full width at 10\% of the maximum. The A and B beams are averaged 
and fit with a transmittance function produced by the SSP to 
refine the wavelength solution. The spectra are then normalized, and 
the spectra of HD~179218 are divided by the spectra of the telluric 
standard whose airmass is scaled in order to minimize the residuals 
of the telluric absorption lines. Regions where the atmospheric transmittance is 
less than 50\% are excluded. The spectra are scaled to flux units 
by adopting the continuum flux provided by the ALLWISE data release, 
which is 4.82$\pm$0.49~Jy \citep{cutri2013}.  

To measure the spectro-astrometric signal of the emission lines, the 
centroids of the A and B rows of the rectified array are measured for 
each column and then averaged. The centroid measurements of the parallel
and anti-parallel position angles are then subtracted and divided by two. 
The spectra and spectroastrometric measurements are presented in Figure 1.

\section{Results}
The SNR of the spectra  
near the center of the order ranges from 100-120 near 2145~cm$^{-1}$ to 
120-140 near 2011~cm$^{-1}$ (Fig 1). The orders at either edge of the 
band have a considerably lower SNR ($\sim$35 near 1915~cm$^{-1}$ 
and 2210 cm$^{-1}$). Moving from the center of the order to the 
edges, the SNR decreases modestly as well. The average SNR of 
the entire spectrum is 55 and is limited by systematic features 
from miscanceled sky lines. 

The FWHM of the PSF ranges from 3.60 pixels in order 102 ($\widetilde{\nu}_0\sim$1976~cm$^{-1}$) 
to 3.46 pixel in order 113 ($\widetilde{\nu}_0\sim$2189~cm$^{-1}$). For 
the plate scale of IRTF in the $M-$band (160~mas/pixel) this corresponds to a FWHM=560~mas. 
There is a weak trend across the orders indicating an improvement in the imaging 
quality of 1.4~mas per order. The 22~mas improvement is comparable to the change in the diffraction limit (420~mas near 1976~cm$^{-1}$ to 380~mas near 2189~cm$^-1$). 

There is a significant fringe running through the data at the 5\% 
level (Fig. 2). This is mostly removed by dividing by the telluric 
standard. There is some residual structure left in each spectrum; 
however, the SNR is dominated in most regions by the quality of 
the telluric correction. 

The iSHELL exposure time calculator (ETC) predicts a SNR of 120/column
for an M=3.95 magnitude star observed with a 0.375$^{\arcsec}$ slit for 
30~min with a 0.56$^{\arcsec}$ PSF. Thus the ETC and delivered
signal to noise are broadly consistent for regions where there is
continuum near the peak of the blaze. 

As noted in section 2, the 
fidelity of the centroid measurement is 
0.4$\times$FWHM/SNR \citep{brannigan2006}. The FWHM of the PSF is 560~mas,
so the fidelity of the centroid measurement at the blue end of the array 
should range from 1.9-2.2~mas near 2145~cm$^{-1}$ and 1.6-1.9~mas near 2011~cm$^{-1}$. 
The measured standard deviations of the centroid measurements near 2145~cm$^{-1}$ and 
2011~cm$^{-1}$ are 2.2~mas and 2.1~mas respectively. The standard deviations of 
the centroid measurements are slightly higher than predicted, pointing to systematic effects. 

The origin of this discrepency is unclear. In principle it could be caused by inconsistencies
in the rectification of the array observed at the parallel slit position 
and the rectification of the array 
observed at the anti-parallel slit position. However, comparison of the rectification of the 
arrays does not show any differences. Another possibility is that 
there are subtle differences in the flat fielding due to different amounts of flexure in the 
system between the observation of the star and the target. Finally, there is some fringing in the 
system. If the shift in the center of the PSF due to the fringes is not repeated exactly, this 
could also result in some remnant structure. To the extent that these effects increase the noise 
of our spectro-astrometric measurement, the effect is to increase the standard deviation of our 
measurement by less than 10\%. The standard deviation of the centroid measurement from 
1950~cm$^{-1}$ - 2200~cm$^{-1}$ is 4.3mas which is consistent with expectations for the average 
SNR over the full array (55). 

The broad wavelength coverage provides simultaneous observation of many CO lines. 
We measured the equivalent width of 155 ro-vibrational CO lines 
spanning v$^{\prime}$=1 through v$^{\prime}$=3 for $^{12}$CO 
and v$^{\prime}$=1 for $^{13}$CO (Table 2). 
Stacking these lines provides enhanced sensitivity of our measurement 
of the line profile and the spectro-astrometric signal (Figs. 3-7). 
The v=1-0 line profile is the average of 48 transitions resulting in 
a SNR along the continuum of 520 and the standard deviation of the
centroid measurement is 0.44mas (Fig. 3). For a system limited by photon noise,
the SNR would improve to 830 and the centroid measurement to 0.28mas. Our results 
do not show this level of improvement because 
no attempt was made to mask the signal from emission not included in the average. 
 
\section{Comparison to CRIRES}
HD~179218 was observed June 16, 2007 with $CRIRES$ on the {\it European
Southern Observatory Very Large Telecope (VLT)} \citep{vdp2015}. CRIRES
is a high resolution (R$\sim$100,000) near infrared spectrograph that 
observes a single order at a time. The 
spectrum covered three regions centered near 2150~cm$^{-1}$, 2105~cm$^{-1}$, 
and 1990~cm$^{-1}$, and the integration times were 20~min, 10~min, 
and 16~min respectively. The spectral grasp of each setting spans 
$\sim$50 cm$^{-1}$ in the $M-$band. The instrument uses four detectors, 
and there is a $\sim$3~cm$^{-1}$ gap between each detector. Two overlapping 
regions of the spectrum of HD~179218 acquired with CRIRES and iSHELL 
are presented in Figure 8. These regions were selected due to the minimal 
telluric lines. The SNR of the continuum from 2009.6~cm$^{-1}$ - 2010.6~cm$^{-1}$ 
is 124 for the CRIRES spectrum which is similar to the iSHELL spectrum 
(122 in 30~minutes; Fig. 8).  Similarly the SNR of the continuum from 2144.4~cm$^{-1}$ to
2145.1~cm$^{-1}$ is 127 for the CRIRES spectrum and 131 for the iSHELL spectrum. 

The large spectral grasp of iSHELL provides a significant advantage 
when stacking multiple transitions. The average profile of the 
v=1--0 CO lines highlights the increase in sensitivity (Fig. 9). The lines averaged
together to create this profile resulted from a total of 46 minutes of integration 
time with CRIRES on the VLT. The average profile of the v=1--0 CO lines constructed from the 
iSHELL spectra required 30 minutes of integration. The improvement in the quality
of the line profile is striking. From the improved line profile, we can place a much 
tighter constraint on the radial distribution of the CO emission. One 
must exercise caution in interpreting the average line profile because
the dense forest of CO lines results in significant blending from v=2--1, 
v=3--2 and $^{13}$CO v=1--0 lines. The impact 
of line blending can be assessed by modeling the spectrum and comparing the average
profile of the observed line and synthetic line. Such modeling is beyond the 
scope of this paper. However to quantify the FWHM of the lines, we fit a flat-topped Gaussian function to each line profile given by

\begin{equation}
    g=a(0)(1-exp(-a(1)e^{-u^2}))+a(4)
\end{equation}

where u = $((v-a(2))/a(3))^3$ and the FWHM is given by 

\begin{equation}
    2\times a(3)\times\sqrt{-ln(ln(2)/a(1))}.
\end{equation} 

This is the functional form for an optically thick emission line, though
here we use it to capture the flat-topped shape of the emission lines.
For each fit, we fixed the line center at 0~km~s$^{-1}$.  From the fits we
find that the FWHM is higher for the hotband transitions than for the
v=1-0 transitions (Figs. 3-7). The fit parameters and FWHM of each average
profile are presented in Table 3. 

\section{Excitation Analysis}
We can estimate the column density and opacity of the CO lines arising in 
the disk around HD~179218. For optically 
thin lines, the number of molecules 
in state $i$ is given by $N_i=L_{ij}/hc\widetilde{\nu}A_{ij}$. 
Assuming the rotational levels of the CO are in local thermodynamic 
equilibrium, the number of emitting molecules in state $i$ is related 
to the total number of molecules in a given vibrational band by

\begin{equation}
    N_i=g_iNe^{-E_i/kT}/Q
\end{equation}

\noindent where $g_i$ is the statistical weight of the upper state, 
$E_i$ is the energy of the upper level, $T$ is the rotational temperature,
and $Q$ is the partition function.  By plotting $ln(N_i/g_i)$ vs $E_i/k$, 
one can determine the rotational temperature from the negative reciprocal 
of the slope and the total number of emitters from the y-intercept (Fig. 10). 
The rotational temperatures of the $^{12}$CO v=1-0, v=2-1, v=3-2, and 
$^{13}$CO v=1-0 lines are 640$\pm$10~K, 770$\pm$30~K, 1270$\pm$120~K, and 630$\pm$90 
respectively. The FWHM of the average profile of the v=1--0 $^{12}$CO and $^{13}$CO 
lines is about 3~km~s$^{-1}$ narrower than the FWHM of the average profile of the v=2--1 and v=3--2
$^{12}$CO lines (Table 3). This is consistent with the higher vibrational
bands originating closer to the star. 

Unlike, van der Plas (2015), we do not detect any emission lines 
from v$^{\prime}$=4 (Fig. 6). The population of the
vibrational states indicates a vibrational temperature of 1850$\pm$90~K. While the 
flux of the v=1-0 emission lines is comparable in our data and the data 
presented by van der Plas et al. (2015), the v=4-3 lines are at least 
4-8 times fainter. This suggests that there is significant temperature 
variability in this system. The fact that the vibrational temperature is greater than the 
rotational temperature of the gas is consistent with UV fluorescence (e.g., 
Brittain et al. 2007), however, it is also possible that the gas is marginally 
optically thick and we are underestimating the rotational temperature. 

The excitation 
diagrams are not linear and are indicative of a large temperature range 
in the emitting region of the disk and/or optically thick ro-vibrational 
lines (Fig. 10). The line center opacity of a line is given by

\begin{equation}
    \tau_i=\Sigma_iA_{ij}(1-e^{-hc\widetilde{\nu}/kT})/8\pi^{3/2}\widetilde{\nu}^3b
\end{equation}

\noindent where $\Sigma_i$ is the column density of CO molecules in 
state $i$, $A_{ij}$ is the Einstein A coefficient for the transition, 
$\widetilde{\nu}$ is the wavenumber of the transition, and $b$ is 
the intrinsic width of the line. We can get a rough idea of the line 
center opacity of the v=1-0 lines by estimating the emitting area of 
the disk from the CO line profile. The half-width at zero intensity 
(HWZI) provides a measure of the inner extent of the CO emission and 
half of the peak separation (HPS) of the line provides a measure of 
the outer extent of the CO emission. The HWZI is $\sim$20km s$^{-1}$ 
and the HPS is $\sim$6km s$^{-1}$. For an inclination of 40$\degr$, 
this translates to an emitting area of $\sim 10^{30}$ cm$^2$. 
At a temperature of 640~K, the most populated state we measure 
(v$^{\prime}$=1, J$^{\prime}$=15) is $3\times10^{40}$molecules. Thus 
the surface density of molecules in that state is $3\times10^{10}$ 
molecules cm$^{-2}$. The line center opacity of that transition is 
$\sim$10$^{-5}$ for $b$=1~km~s$^{-1}$. Thus the lines are extremely 
optically thin. This is expected for lines populated by UV fluorescence (Brittain et al. 2009). 
A more quantitative measure of the CO distribution and abundance requires
spectral synthesis of the spectrum and fitting to the SA signal (e.g., Brittain et al. 2007), which 
is the subject of future work.

\section{conclusions}
We have measured the fundamental ro-vibrational CO spectrum 
from the transition disk of HD~179218 using iSHELL on the 
IRTF. The lines are resolved, symmetric, and consistent with the line 
profile observed with CRIRES in 2007. We have also measured 
the spectro-astrometric signal of the CO 
lines and the signal is qualitatively
consistent with gas emission in Keplerian orbit. 
In a future paper we will 
compare this source to other measurements 
of CO emission in transition disks and 
model the spectro-astrometric measurement and emission lines spectrum.

This work highlights the utility of iSHELL for $M-$band 
spectroscopy of young stars and for spectroastrometry. 
Sub-milliarcsecond fidelity is achievable with this instrument 
on the IRTF. The ETC provides a reliable estimate of the instrument 
sensitivity, though there is significant variation across the 
orders. The comparison of iSHELL and CRIRES highlights the high 
science value of modest aperture 3m class telescopes when equipped 
with 21$^{\rm st}$ century instrumentation. When the scientific objective depends on the 
observation of multiple lines spanning the photometric band, 
the much larger spectral grasp of iSHELL enhances its efficiency. 
While an updated CRIRES with a cross-dispersed spectrograph 
promises similar gains in efficiency, iSHELL will remain a 
valuable counterpart in the northern hemisphere.

\acknowledgments
SDB, JSC, and JRN were visiting Astronomers at the Infrared Telescope Facility, which is operated by the University of Hawaii under contract NNH14CK55B with the National Aeronautics and Space Administration. Work by SDB was performed in part at the National Optical Astronomy Observatory. NOAO is operated by the Association of Universities for Research in Astronomy (AURA), Inc. under a cooperative agreement with the National Science Foundation. SDB also acknowledges support from this work by NASA Agreement No. NXX15AD94G; NASA Agreement No. NNX16AJ81G; and NSF-AST 1517014. JRN acknowledges the stimulating research environment supported by NASA Agreement No. NNX15AD94G to the "Earths in Other Solar Systems" program. Basic research in infrared astronomy at the Naval Research Laboratory is supported by 6.1 base funding. This work has made use of data from the European Space Agency
mission {\it Gaia} (\url{https://www.cosmos.esa.int/gaia}), processed by the {\it Gaia} Data Processing and Analysis Consortium (DPAC,
\url{https://www.cosmos.esa.int/web/gaia/dpac/consortium}). Funding
for the DPAC has been provided by national institutions, in particular
the institutions participating in the {\it Gaia} Multilateral Agreement.

{\it Facilities:} \facility{Infrared Telescope Facility (iSHELL)}

\nocite{*}
\bibliographystyle{apj}
\bibliography{ms}		

\begin{deluxetable*}{cccc}
\tabletypesize{\scriptsize}
\tablecaption{Journal of Observations}
\tablewidth{0pt}
\tablehead{ \colhead{Star}  & \colhead{M$_{\rm m}$} & \colhead{Int Time} & \colhead{PA} \\
 {} & {} & \colhead{(m)} & {}  }
\startdata
HD~177724 & 2.97 & 15 & 75.5$\degr$ \\
HD~179218 & 3.87 & 15 & 23$\degr$ \\
HD~179218 & 3.87 & 15 & 203$\degr$ \\ 
\enddata

\label{tab:Md}
\end{deluxetable*}

\startlongtable
\begin{deluxetable*}{ccccccc}
\tabletypesize{\scriptsize}
\tablecaption{Observed CO lines}
\tablewidth{0pt}
\tablehead{
\colhead{v$^{\prime}$} & \colhead{v$^{\prime\prime}$} & ID & $\widetilde{\nu}$                  & \colhead{A}	         & \colhead{E$^{\prime}$} & \colhead{EqW} \\
 	                   &		                      &	   &	\colhead{cm$^{-1}$}	    &	\colhead{s$^{-1}$}	 &	\colhead{cm$^{-1}$}	  &	\colhead{cm$^{-1}$} }
\startdata	
 \multicolumn{7}{c}{ $^{12}$C$^{16}$O~lines}\\
 \hline
1	&	0	&	R(14)	&	2196.66	&	18.3	&	3681.71	&	0.0141 $\pm$ 0.0010	\\
1	&	0	&	R(13)	&	2193.36	&	18.1	&	3624.64	&	0.0154	$\pm$	0.0011	\\
1	&	0	&	R(12)	&	2190.02	&	18.0	&	3571.37	&	0.0131	$\pm$	0.0004	\\
1	&	0	&	R(11)	&	2186.64	&	17.9	&	3521.89	&	0.0164	$\pm$	0.0013	\\
1	&	0	&	R(10)	&	2183.22	&	17.7	&	3476.21	&	0.0120	$\pm$	0.0004	\\
1	&	0	&	R(8)	&	2176.28	&	17.4	&	3396.26	&	0.0138	$\pm$	0.0007	\\
1	&	0	&	R(7)	&	2172.76	&	17.2	&	3361.99	&	0.0143	$\pm$	0.0007	\\
1	&	0	&	R(6)	&	2169.20	&	16.9	&	3331.52	&	0.0126	$\pm$	0.0006	\\
1	&	0	&	R(5)	&	2165.60	&	16.6	&	3304.86	&	0.0147	$\pm$	0.0004	\\
1	&	0	&	R(4)	&	2161.97	&	16.3	&	3282.00	&	0.0147	$\pm$	0.0073	\\
1	&	0	&	R(3)	&	2158.30	&	15.9	&	3262.96	&	0.0146	$\pm$	0.0004	\\
1	&	0	&	R(2)	&	2154.60	&	15.2	&	3247.72	&	0.0100	$\pm$	0.0003	\\
1	&	0	&	R(1)	&	2150.86	&	14.1	&	3236.29	&	0.0133	$\pm$	0.0004	\\
1	&	0	&	R(0)	&	2147.08	&	11.7	&	3228.67	&	0.0088	$\pm$	0.0005	\\
1	&	0	&	P(1)	&	2139.43	&	34.7	&	3224.86	&	0.0090	$\pm$	0.0010	\\
1	&	0	&	P(2)	&	2135.55	&	23.0	&	3228.67	&	0.0070	$\pm$	0.0020	\\
1	&	0	&	P(3)	&	2131.63	&	20.6	&	3236.29	&	0.0130	$\pm$	0.0013	\\
1	&	0	&	P(4)	&	2127.68	&	19.5	&	3247.71	&	0.0176	$\pm$	0.0005	\\
1	&	0	&	P(5)	&	2123.70	&	18.9	&	3262.96	&	0.0135	$\pm$	0.0004	\\
1	&	0	&	P(6)	&	2119.68	&	18.4	&	3282.00	&	0.0135	$\pm$	0.0007	\\
1	&	0	&	P(7)	&	2115.63	&	18.1	&	3304.86	&	0.0144	$\pm$	0.0014	\\
1	&	0	&	P(8)	&	2111.54	&	17.8	&	3331.52	&	0.0117	$\pm$	0.0004	\\
1	&	0	&	P(9)	&	2107.42	&	17.6	&	3361.98	&	0.0150	$\pm$	0.0030	\\
1	&	0	&	P(10)	&	2103.27	&	17.3	&	3396.26	&	0.0171	$\pm$	0.0034	\\
1	&	0	&	P(11)	&	2099.08	&	17.2	&	3434.33	&	0.0145	$\pm$	0.0004	\\
1	&	0	&	P(12)	&	2094.86	&	17.0	&	3476.21	&	0.0114	$\pm$	0.0011	\\
1	&	0	&	P(13)	&	2090.61	&	16.8	&	3521.89	&	0.0136	$\pm$	0.0027	\\
1	&	0	&	P(14)	&	2086.32	&	16.7	&	3571.37	&	0.0108	$\pm$	0.0011	\\
1	&	0	&	P(15)	&	2082.00	&	16.5	&	3624.64	&	0.0146	$\pm$	0.0029	\\
1	&	0	&	P(17)	&	2073.26	&	16.3	&	3742.57	&	0.0114	$\pm$	0.0006	\\
1	&	0	&	P(18)	&	2068.85	&	16.1	&	3807.22	&	0.0174	$\pm$	0.0035	\\
1	&	0	&	P(20)	&	2059.91	&	15.9	&	3947.88	&	0.0158	$\pm$	0.0047	\\
1	&	0	&	P(25)	&	2037.03	&	15.3	&	4365.67	&	0.0084	$\pm$	0.0008	\\
1	&	0	&	P(26)	&	2032.35	&	15.2	&	4460.54	&	0.0043	$\pm$	0.0001	\\
1	&	0	&	P(27)	&	2027.65	&	15.1	&	4559.17	&	0.0056	$\pm$	0.0006	\\
1	&	0	&	P(28)	&	2022.91	&	14.9	&	4661.56	&	0.0056	$\pm$	0.0011	\\
1	&	0	&	P(29)	&	2018.15	&	14.8	&	4767.71	&	0.0081	$\pm$	0.0008	\\
1	&	0	&	P(30)	&	2013.35	&	14.7	&	4877.60	&	0.0039	$\pm$	0.0012	\\
1	&	0	&	P(31)	&	2008.53	&	14.6	&	4991.25	&	0.0058	$\pm$	0.0017	\\
1	&	0	&	P(32)	&	2003.67	&	14.5	&	5108.63	&	0.0011	$\pm$	0.0010	\\
2	&	1	&	R(22)	&	2194.46	&	36.8	&	6381.69	&	0.0061	$\pm$	0.0012	\\
2	&	1	&	R(21)	&	2191.50	&	36.6	&	6295.17	&	0.0053	$\pm$	0.0011	\\
2	&	1	&	R(19)	&	2185.45	&	36.2	&	6133.33	&	0.0069	$\pm$	0.0014	\\
2	&	1	&	R(18)	&	2182.36	&	36.0	&	6058.02	&	0.0013	$\pm$	0.0013	\\
2	&	1	&	R(14)	&	2169.66	&	35.1	&	5794.30	&	0.0048	$\pm$	0.0010	\\
2	&	1	&	R(13)	&	2166.39	&	34.8	&	5737.76	&	0.0061	$\pm$	0.0006	\\
2	&	1	&	R(12)	&	2163.08	&	34.6	&	5684.97	&	0.0050	$\pm$	0.0015	\\
2	&	1	&	R(10)	&	2156.36	&	34.0	&	5590.70	&	0.0053	$\pm$	0.0005	\\
2	&	1	&	R(8)	&	2149.49	&	33.3	&	5511.48	&	0.0040	$\pm$	0.0013	\\
2	&	1	&	R(7)	&	2146.00	&	32.9	&	5477.52	&	0.0067	$\pm$	0.0013	\\
2	&	1	&	R(6)	&	2142.47	&	32.5	&	5447.33	&	0.0013	$\pm$	0.0013	\\
2	&	1	&	R(5)	&	2138.91	&	32.0	&	5420.91	&	0.0037	$\pm$	0.0010	\\
2	&	1	&	R(4)	&	2135.31	&	31.3	&	5398.26	&	0.0032	$\pm$	0.0003	\\
2	&	1	&	R(0)	&	2120.57	&	22.5	&	5345.43	&	0.0020	$\pm$	0.0010	\\
2	&	1	&	P(1)	&	2112.98	&	66.7	&	5341.65	&	0.0024	$\pm$	0.0013	\\
2	&	1	&	P(2)	&	2109.14	&	44.2	&	5345.43	&	0.0031	$\pm$	0.0006	\\
2	&	1	&	P(3)	&	2105.26	&	39.6	&	5352.98	&	0.0028	$\pm$	0.0006	\\
2	&	1	&	P(4)	&	2101.34	&	37.5	&	5364.29	&	0.0012	$\pm$	0.0013	\\
2	&	1	&	P(5)	&	2097.39	&	36.2	&	5379.39	&	0.0030	$\pm$	0.0009	\\
2	&	1	&	P(6)	&	2093.41	&	35.4	&	5398.27	&	0.0022	$\pm$	0.0010	\\
2	&	1	&	P(7)	&	2089.39	&	34.7	&	5420.91	&	0.0091	$\pm$	0.0005	\\
2	&	1	&	P(8)	&	2085.34	&	34.2	&	5447.33	&	0.0024	$\pm$	0.0010	\\
2	&	1	&	P(11)	&	2072.99	&	33.0	&	5549.20	&	0.0035	$\pm$	0.0001	\\
2	&	1	&	P(14)	&	2060.33	&	32.0	&	5684.97	&	0.0046	$\pm$	0.0014	\\
2	&	1	&	P(15)	&	2056.05	&	31.8	&	5737.76	&	0.0059	$\pm$	0.0002	\\
2	&	1	&	P(16)	&	2051.73	&	31.5	&	5794.30	&	0.0039	$\pm$	0.0010	\\
2	&	1	&	P(17)	&	2047.38	&	31.2	&	5854.60	&	0.0037	$\pm$	0.0001	\\
2	&	1	&	P(18)	&	2043.00	&	31.0	&	5918.66	&	0.0022	$\pm$	0.0007	\\
2	&	1	&	P(19)	&	2038.58	&	30.7	&	5986.46	&	0.0045	$\pm$	0.0005	\\
2	&	1	&	P(20)	&	2034.14	&	30.5	&	6058.03	&	0.0050	$\pm$	0.0005	\\
2	&	1	&	P(21)	&	2029.66	&	30.3	&	6133.33	&	0.0031	$\pm$	0.0005	\\
2	&	1	&	P(23)	&	2020.60	&	29.8	&	6295.17	&	0.0010	$\pm$	0.0010	\\
2	&	1	&	P(24)	&	2016.03	&	29.6	&	6381.70	&	0.0022	$\pm$	0.0006	\\
2	&	1	&	P(25)	&	2011.42	&	29.4	&	6471.96	&	0.0023	$\pm$	0.0006	\\
2	&	1	&	P(27)	&	2002.12	&	28.9	&	6663.68	&	0.0010	$\pm$	0.0010	\\
3	&	2	&	R(30)	&	2189.31	&	55.0	&	9281.07	&	0.0027	$\pm$	0.0008	\\
3	&	2	&	R(27)	&	2181.33	&	54.3	&	8946.45	&	0.0020	$\pm$	0.0010	\\
3	&	2	&	R(26)	&	2178.59	&	54.0	&	8842.27	&	0.0051	$\pm$	0.0010	\\
3	&	2	&	R(25)	&	2175.81	&	53.7	&	8741.76	&	0.0010	$\pm$	0.0010	\\
3	&	2	&	R(23)	&	2170.14	&	53.2	&	8551.84	&	0.0032	$\pm$	0.0005	\\
3	&	2	&	R(22)	&	2167.25	&	52.9	&	8462.42	&	0.0020	$\pm$	0.0005	\\
3	&	2	&	R(21)	&	2164.31	&	52.6	&	8376.69	&	0.0030	$\pm$	0.0005	\\
3	&	2	&	R(20)	&	2161.34	&	52.3	&	8294.67	&	0.0024	$\pm$	0.0005	\\
3	&	2	&	R(18)	&	2155.28	&	51.8	&	8141.75	&	0.0029	$\pm$	0.0005	\\
3	&	2	&	R(17)	&	2152.20	&	51.4	&	8070.86	&	0.0015	$\pm$	0.0005	\\
3	&	2	&	R(16)	&	2149.08	&	51.1	&	8003.68	&	0.0017	$\pm$	0.0005	\\
3	&	2	&	R(14)	&	2142.72	&	50.5	&	7880.48	&	0.0010	$\pm$	0.0005	\\
3	&	2	&	R(12)	&	2136.21	&	49.7	&	7772.16	&	0.0006	$\pm$	0.0005	\\
3	&	2	&	R(11)	&	2132.91	&	49.3	&	7723.6	&	0.0014	$\pm$	0.0005	\\
3	&	2	&	R(10)	&	2129.56	&	48.9	&	7678.76	&	0.0028	$\pm$	0.0005	\\
3	&	2	&	R(9)	&	2126.18	&	48.5	&	7637.66	&	0.0026	$\pm$	0.0010	\\
3	&	2	&	R(8)	&	2122.76	&	48.0	&	7600.28	&	0.0020	$\pm$	0.0010	\\
3	&	2	&	R(7)	&	2119.30	&	47.4	&	7566.63	&	0.0018	$\pm$	0.0010	\\
3	&	2	&	R(5)	&	2112.29	&	46.0	&	7510.56	&	0.0018	$\pm$	0.0010	\\
3	&	2	&	R(4)	&	2108.72	&	45.1	&	7488.12	&	0.0034	$\pm$	0.0005	\\
3	&	2	&	R(3)	&	2105.13	&	43.8	&	7469.43	&	0.0008	$\pm$	0.0010	\\
3	&	2	&	R(1)	&	2097.82	&	39.0	&	7443.24	&	0.0020	$\pm$	0.0010	\\
3	&	2	&	P(4)	&	2075.07	&	54.0	&	7454.47	&	0.0035	$\pm$	0.0010	\\
3	&	2	&	P(5)	&	2071.15	&	52.2	&	7469.42	&	0.0016	$\pm$	0.0010	\\
3	&	2	&	P(6)	&	2067.21	&	50.9	&	7488.12	&	0.0019	$\pm$	0.0010	\\
3	&	2	&	P(8)	&	2059.21	&	49.2	&	7536.73	&	0.0016	$\pm$	0.0010	\\
3	&	2	&	P(9)	&	2055.16	&	48.6	&	7566.64	&	0.0038	$\pm$	0.0010	\\
3	&	2	&	P(10)	&	2051.08	&	48.0	&	7600.28	&	0.0023	$\pm$	0.0010	\\
3	&	2	&	P(11)	&	2046.96	&	47.5	&	7637.65	&	0.0016	$\pm$	0.0010	\\
3	&	2	&	P(12)	&	2042.81	&	47.0	&	7678.76	&	0.0020	$\pm$	0.0020	\\
3	&	2	&	P(15)	&	2030.16	&	45.8	&	7824.46	&	0.0026	$\pm$	0.0005	\\
4	&	3	&	R(32)	&	2166.94	&	70.9	&	11566.89	&	0.0010	$\pm$	0.0010	\\
4	&	3	&	R(31)	&	2164.43	&	70.6	&	11445.50	&	0.0010	$\pm$	0.0010	\\
4	&	3	&	R(29)	&	2159.30	&	69.9	&	11213.62	&	0.0010	$\pm$	0.0010	\\
4	&	3	&	R(28)	&	2156.67	&	69.6	&	11103.12	&	0.0010	$\pm$	0.0010	\\
4	&	3	&	R(27)	&	2154.01	&	69.3	&	10996.28	&	0.0015	$\pm$	0.0010	\\
4	&	3	&	R(25)	&	2148.56	&	68.6	&	10793.52	&	0.0010	$\pm$	0.0010	\\
4	&	3	&	R(23)	&	2142.95	&	68.0	&	10605.36	&	0.0010	$\pm$	0.0010	\\
4	&	3	&	R(22)	&	2140.09	&	67.6	&	10516.78	&	0.0026	$\pm$	0.0010	\\
4	&	3	&	R(21)	&	2137.20	&	67.3	&	10431.87	&	0.0010	$\pm$	0.0010	\\
4	&	3	&	R(19)	&	2131.29	&	66.5	&	10273.04	&	0.0010	$\pm$	0.0010	\\
4	&	3	&	R(18)	&	2128.27	&	66.1	&	10199.13	&	0.0020	$\pm$	0.0010	\\
4	&	3	&	R(15)	&	2119.01	&	64.9	&	9999.49	&	0.0010	$\pm$	0.0010	\\
4	&	3	&	R(13)	&	2112.65	&	64.0	&	9884.82	&	0.0011	$\pm$	0.0010	\\
4	&	3	&	R(7)	&	2092.68	&	60.6	&	9629.41	&	0.0010	$\pm$	0.0010	\\
4	&	3	&	P(2)	&	2056.52	&	81.4	&	9499.76	&	0.0010	$\pm$	0.0010	\\
4	&	3	&	P(3)	&	2052.71	&	72.9	&	9507.17	&	0.0010	$\pm$	0.0010	\\
4	&	3	&	P(4)	&	2048.86	&	69.0	&	9518.28	&	0.0010	$\pm$	0.0010	\\
4	&	3	&	P(6)	&	2041.07	&	65.1	&	9551.62	&	0.0025	$\pm$	0.0010	\\
4	&	3	&	P(8)	&	2033.14	&	63.0	&	9599.77	&	0.0018	$\pm$	0.0010	\\
4	&	3	&	P(9)	&	2029.13	&	62.1	&	9629.41	&	0.0016	$\pm$	0.0010	\\
4	&	3	&	P(11)	&	2021.00	&	60.7	&	9699.76	&	0.0010	$\pm$	0.0010	\\
4	&	3	&	P(15)	&	2004.34	&	58.5	&	9884.82	&	0.0010	$\pm$	0.0010	\\
4	&	3	&	P(16)	&	2000.09	&	58.1	&	9940.31	&	0.0019	$\pm$	0.0010	\\
 \hline
 \multicolumn{7}{c}{ $^{13}$C$^{16}$O~lines}\\
 \hline
1	&	0	&	R(30)	&	2193.13	&	18.2	&	4955.16	&	0.0010	$\pm$	0.0010	\\
1	&	0	&	R(28)	&	2187.89	&	18.0	&	4734.18	&	0.0010	$\pm$	0.0010	\\
1	&	0	&	R(23)	&	2174.17	&	17.6	&	4244.57	&	0.0019	$\pm$	0.0010	\\
1	&	0	&	R(22)	&	2171.31	&	17.5	&	4157.44	&	0.0014	$\pm$	0.0010	\\
1	&	0	&	R(21)	&	2168.42	&	17.4	&	4073.91	&	0.0010	$\pm$	0.0010	\\
1	&	0	&	R(16)	&	2153.44	&	16.9	&	3710.50	&	0.0013	$\pm$	0.0010	\\
1	&	0	&	R(13)	&	2144.03	&	16.6	&	3535.90	&	0.0027	$\pm$	0.0005	\\
1	&	0	&	R(10)	&	2134.31	&	16.2	&	3393.96	&	0.0046	$\pm$	0.0005	\\
1	&	0	&	R(9)	&	2131.00	&	16.0	&	3353.92	&	0.0032	$\pm$	0.0005	\\
1	&	0	&	R(7)	&	2124.29	&	15.7	&	3284.74	&	0.0017	$\pm$	0.0010	\\
1	&	0	&	R(5)	&	2117.43	&	15.2	&	3230.11	&	0.0023	$\pm$	0.0010	\\
1	&	0	&	R(4)	&	2113.95	&	14.9	&	3208.25	&	0.0026	$\pm$	0.0010	\\
1	&	0	&	R(3)	&	2110.44	&	14.5	&	3190.04	&	0.0017	$\pm$	0.0010	\\
1	&	0	&	R(2)	&	2106.90	&	13.9	&	3175.47	&	0.0011	$\pm$	0.0010	\\
1	&	0	&	P(1)	&	2092.39	&	31.8	&	3153.61	&	0.0013	$\pm$	0.0010	\\
1	&	0	&	P(3)	&	2084.94	&	18.8	&	3164.54	&	0.0025	$\pm$	0.0010	\\
1	&	0	&	P(7)	&	2069.66	&	16.5	&	3230.11	&	0.0042	$\pm$	0.0010	\\
1	&	0	&	P(9)	&	2061.82	&	16.1	&	3284.74	&	0.0038	$\pm$	0.0020	\\
1	&	0	&	P(10)	&	2057.86	&	15.9	&	3317.51	&	0.0026	$\pm$	0.0010	\\
1	&	0	&	P(12)	&	2049.83	&	15.6	&	3393.96	&	0.0028	$\pm$	0.0010	\\
1	&	0	&	P(13)	&	2045.78	&	15.4	&	3437.65	&	0.0035	$\pm$	0.0005	\\
1	&	0	&	P(15)	&	2037.57	&	15.2	&	3535.91	&	0.0038	$\pm$	0.0010	\\
1	&	0	&	P(16)	&	2033.42	&	15.0	&	3590.48	&	0.0028	$\pm$	0.0010	\\
1	&	0	&	P(17)	&	2029.24	&	14.9	&	3648.68	&	0.0025	$\pm$	0.0005	\\
1	&	0	&	P(19)	&	2020.78	&	14.7	&	3775.95	&	0.0033	$\pm$	0.0010	\\
1	&	0	&	P(21)	&	2012.21	&	14.5	&	3917.70	&	0.0018	$\pm$	0.0010	\\
\enddata
\label{tab:EqWs}
\end{deluxetable*}

\begin{deluxetable*}{ccccccc}
\tabletypesize{\scriptsize}
\tablecaption{Line Profile Fit Parameters}
\tablewidth{0pt}
\tablehead{ \colhead{Profile}  & \colhead{a(0)} & \colhead{a(1)}  & \colhead{a(2)}& \colhead{a(3)} & \colhead{a(4)}& \colhead{FWHM} \\
 {} & {} & {} & {}& {}& {}& {km s$^{-1}$} \\ }
\startdata
$^{12}$CO v=1--0 & 0.082$\pm$0.001 & 4.4$\pm$0.4 & 0 & 7.2$\pm$0.1 &1.0024$\pm$0.0001 & 19.51$\pm$0.27 \\
$^{12}$CO v=2--1 & 0.021$\pm$0.001 & 30.1$\pm$9.1 & 0 & 5.8$\pm$0.5 & 1.0046$\pm$0.0001 & 22.64$\pm$0.66 \\
$^{12}$CO v=3--2 & 0.013$\pm$0.003 & 32.1$\pm$17.5 & 0 & 6$\pm$4 & 1.0002$\pm$0.0001 & 22.17$\pm$1.07 \\
$^{12}$CO v=4--3 & 0.004$\pm$0.003 & 200$\pm$200 & 0 & 5$\pm$7 & 1.005$\pm$0.001 & \nodata \\
$^{13}$CO v=1--0 & 0.016$\pm$0.001 & 8$\pm$3 & 0 & 6.4$\pm$0.9 & 1.005$\pm$0.0001 & 20.15$\pm$0.91 \\
\enddata
\label{tab:Md}
\end{deluxetable*}

\begin{figure*}
        \includegraphics[width=0.95\textwidth]{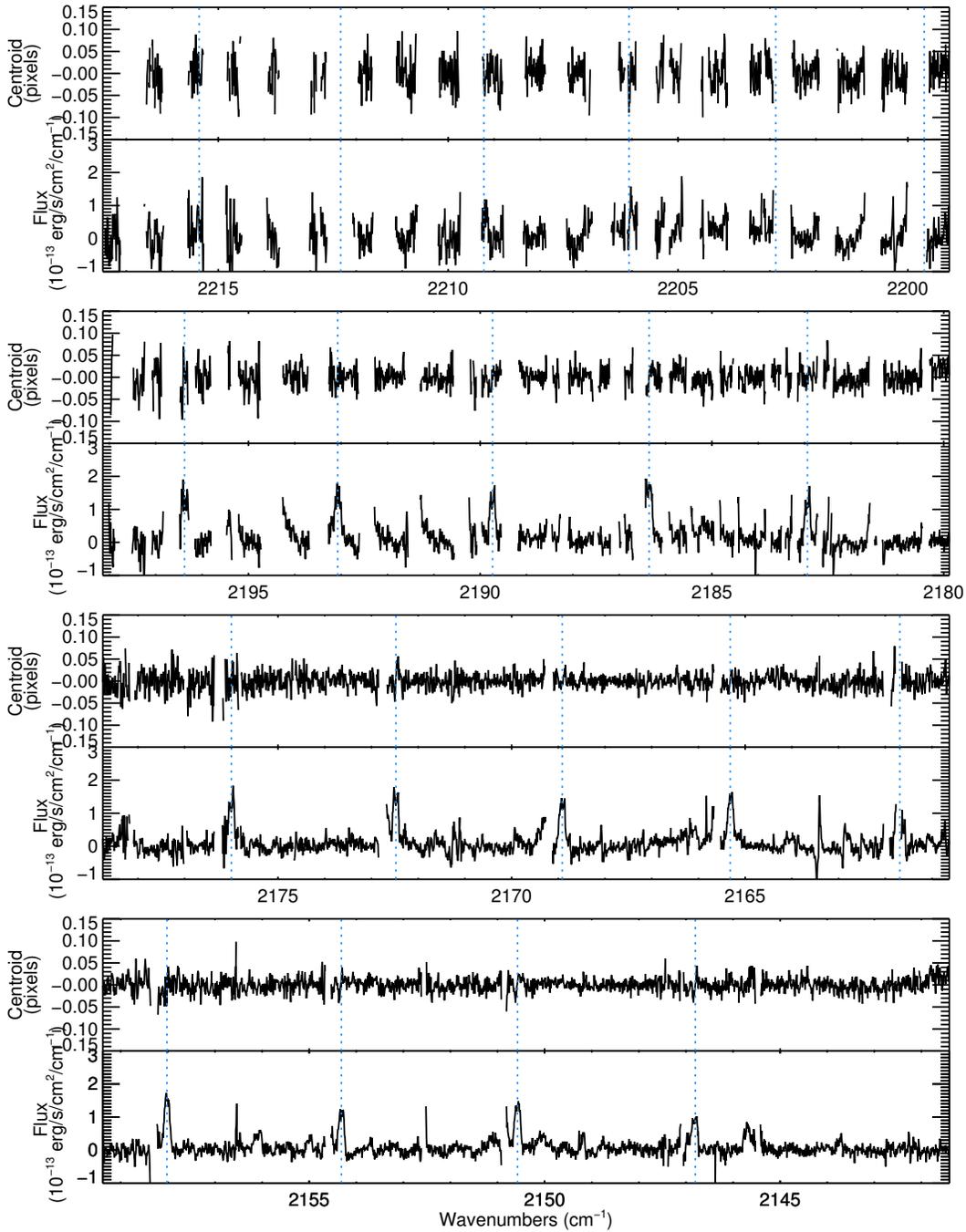}
        \caption{$M-$band spectrum of HD~179218. The spectrum of HD~179218 is plotted in the lower panel and the relative centroid of the PSF is plotted in the upper panel. Gaps in the spectra are regions where the atmospheric transmittance is less than 50\%.  The positions of the v=1-0 CO lines are marked with dotted vertical lines (blue).   }
        \label{fig:1a}
\end{figure*}

\setcounter{figure}{1}
\begin{figure*}\ContinuedFloat
        \includegraphics[width=0.95\textwidth]{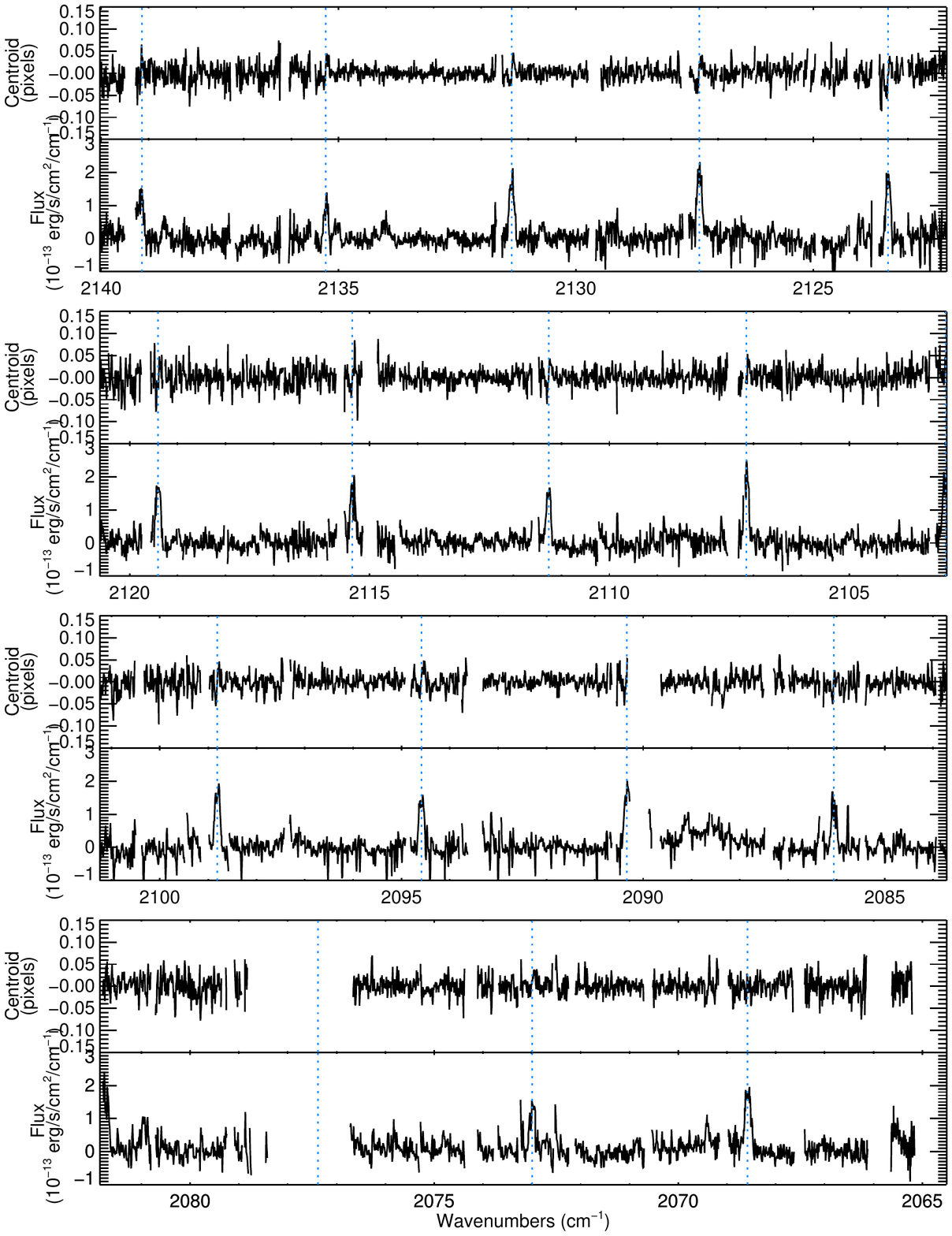}
        \caption{cont}
        \label{fig:1b}
\end{figure*}

\setcounter{figure}{1}
\begin{figure*}\ContinuedFloat
        \includegraphics[width=0.95\textwidth]{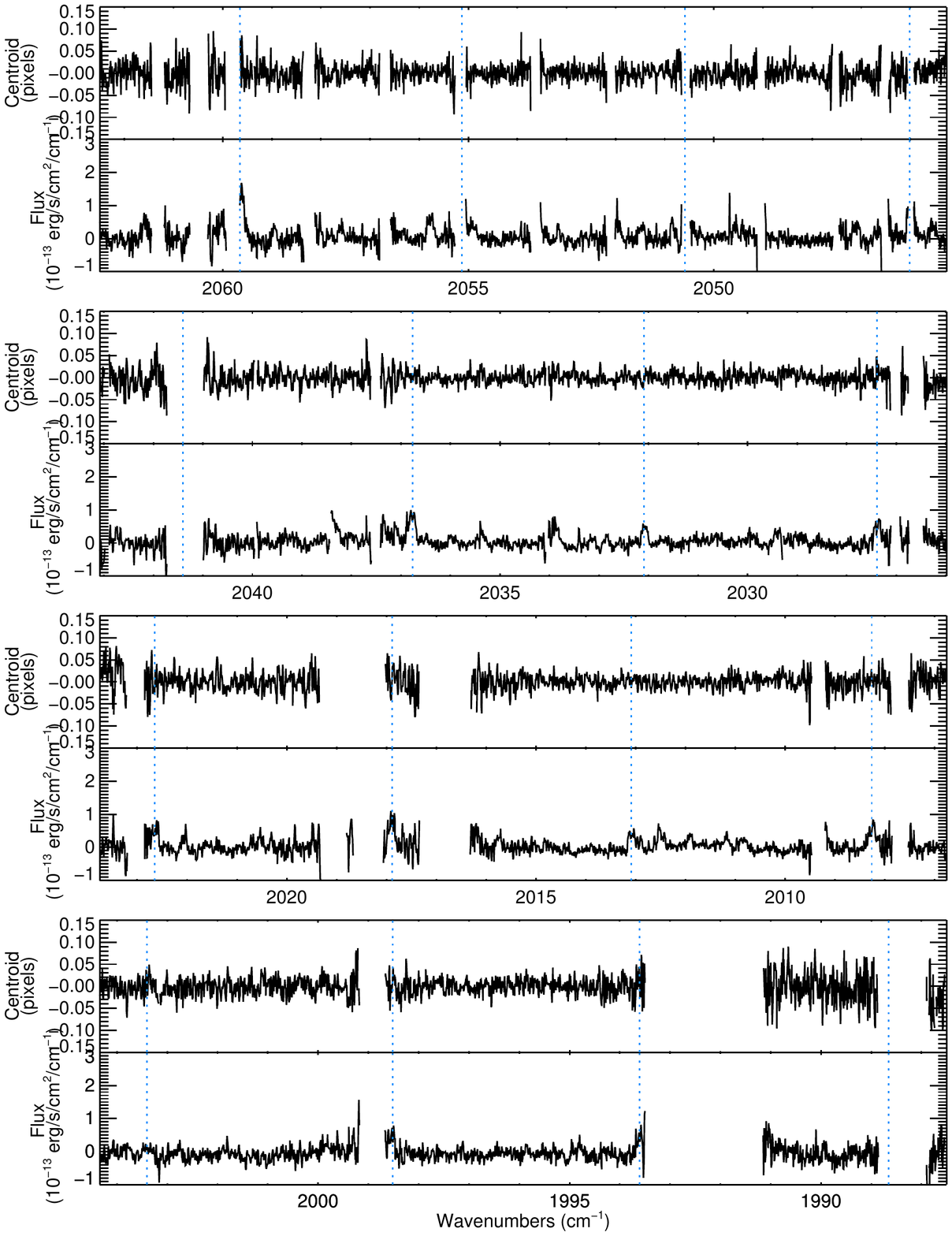}
        \caption{cont}
        \label{fig:1c}
\end{figure*}

\setcounter{figure}{1}
\begin{figure*}\ContinuedFloat
        \includegraphics[width=0.95\textwidth]{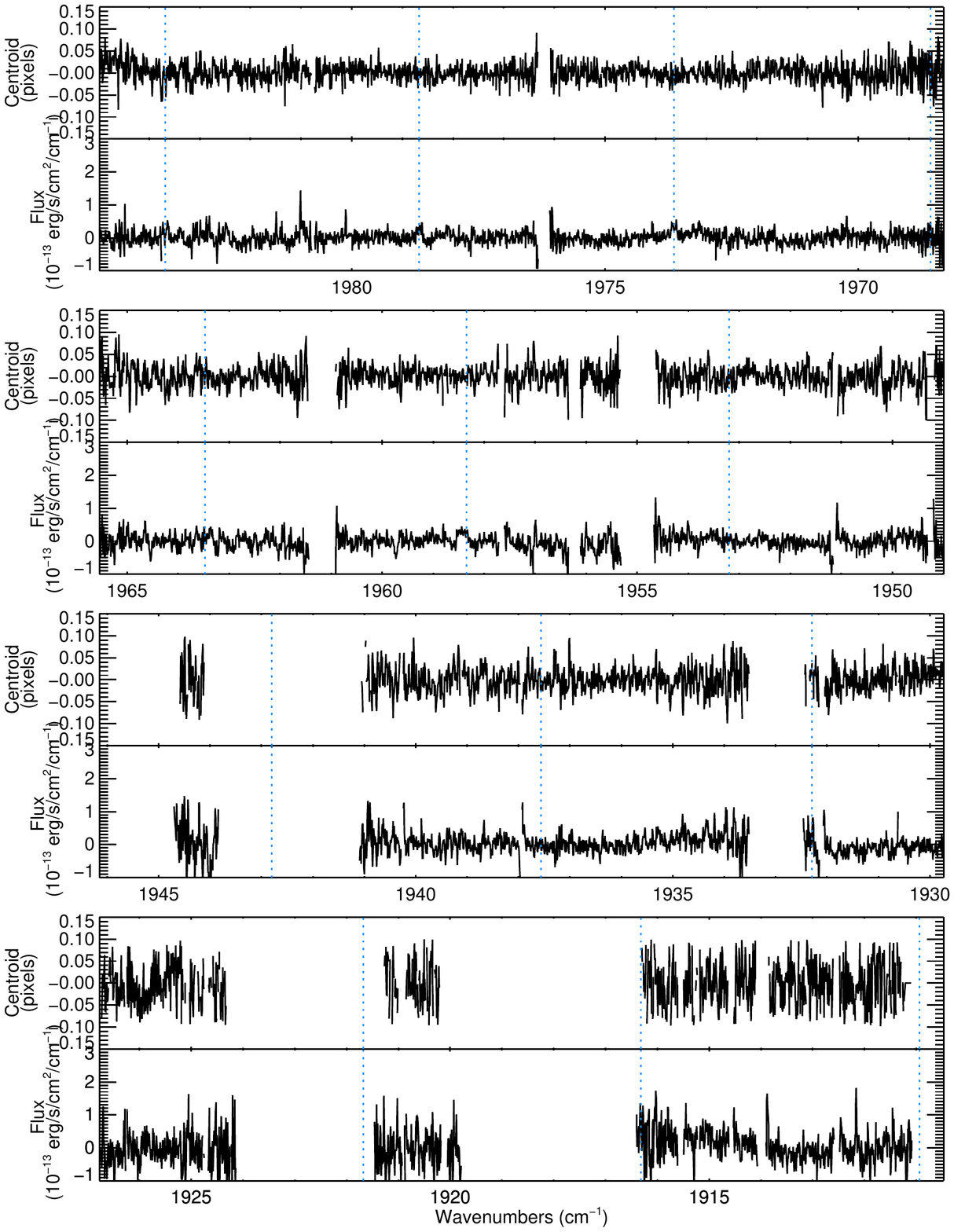}
        \caption{cont}
        \label{fig:1d}
\end{figure*}

\begin{figure*}
        \includegraphics[width=0.95\textwidth]{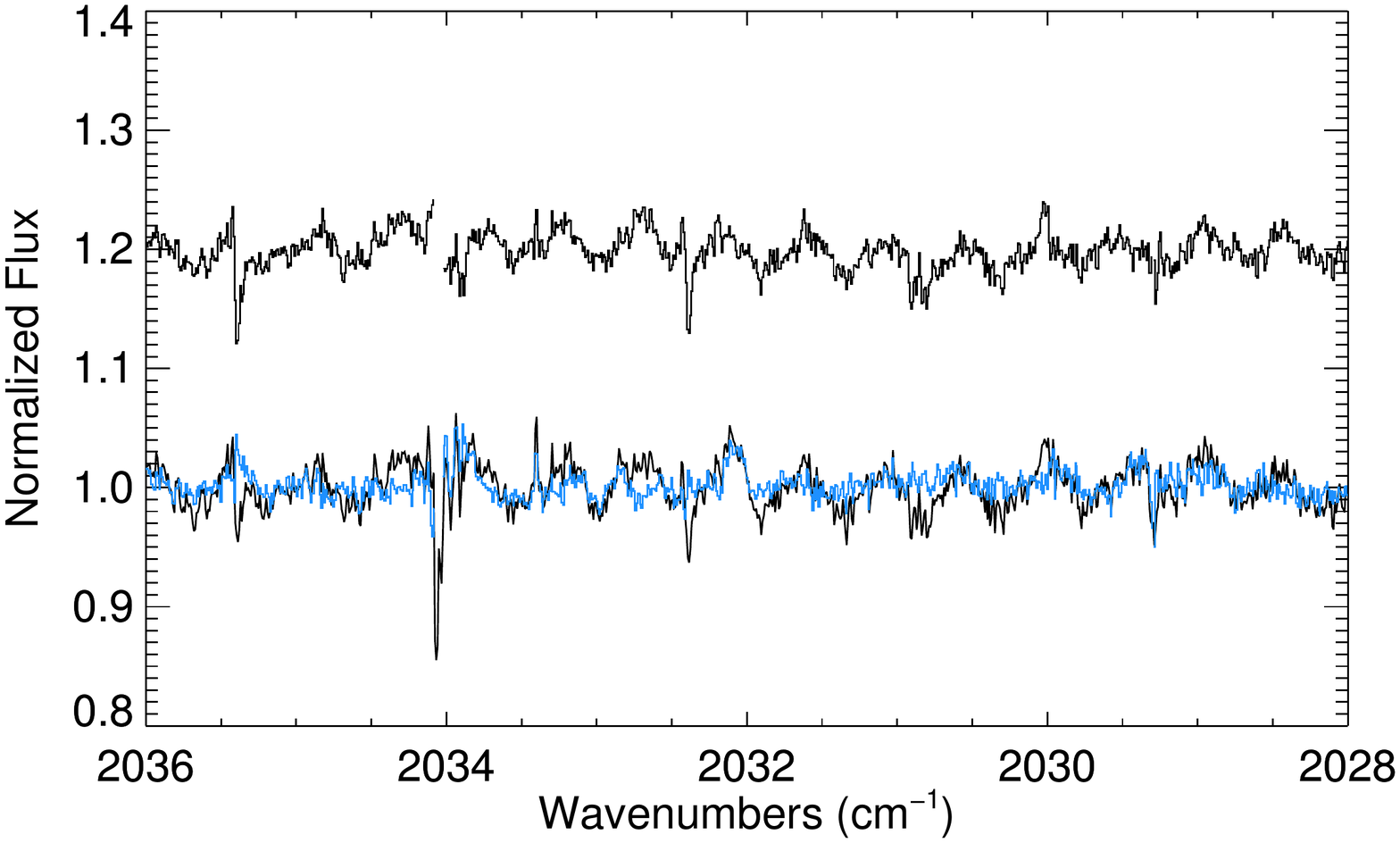}
        \caption{Sample spectrum of HD~179218 illustrating fringing. This region was selected due to the minimal telluric features. Rather than dividing by the telluric standard, the atmospheric lines were corrected with the SSP model (black line). The fringe is $\sim$5\% of the continuum in this region which is representative of the spectrum. Telluric correction with a standard star taken without having moved the grating between observations corrects the fringe (blue line). The difference between the two versions is plotted above.}
        \label{fig:2}
\end{figure*}

\begin{figure*}
        \includegraphics[width=0.9\textwidth]{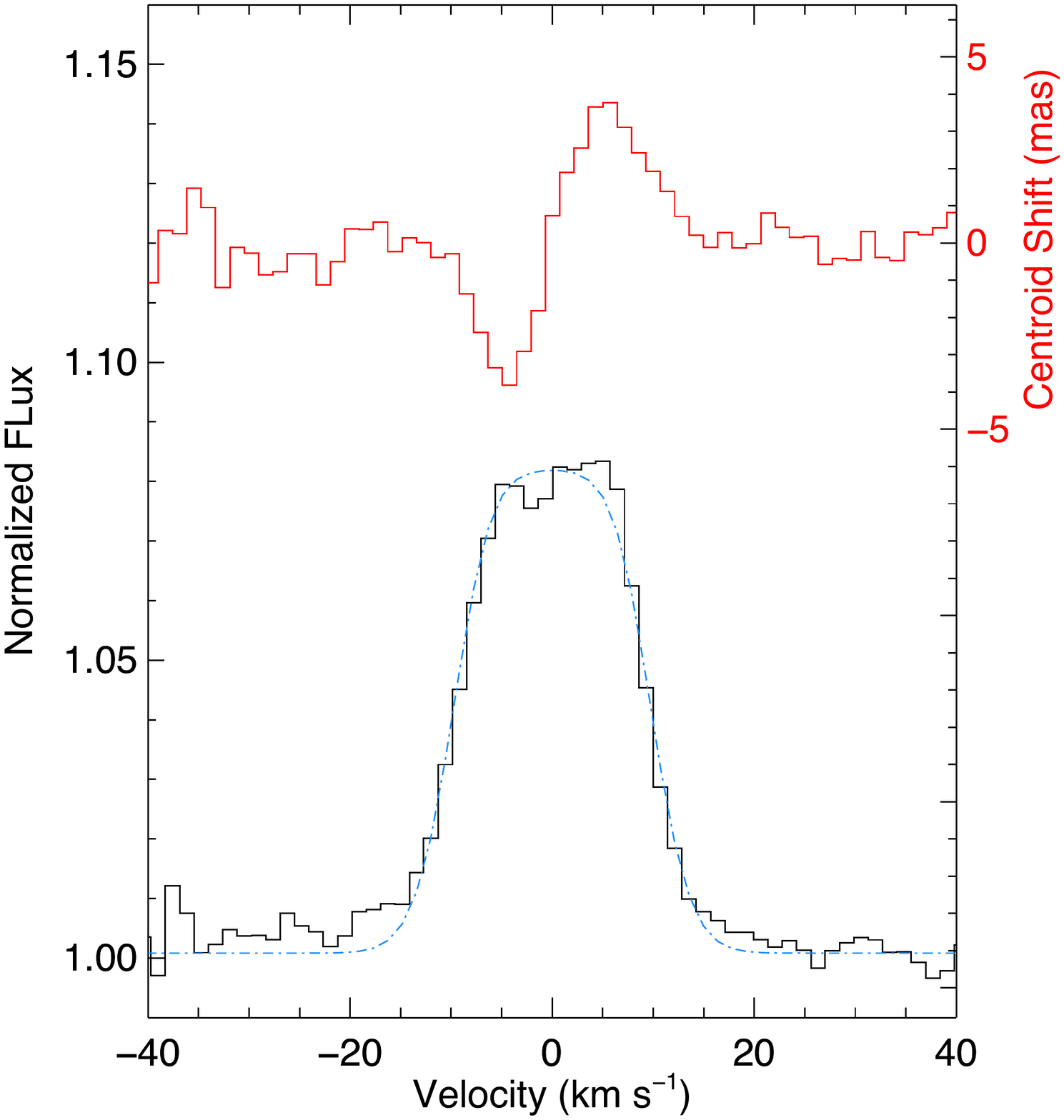}
        \caption{Profile of v=1--0 CO lines and spectroastrometric signal. The average profile of the v=1--0 spectral lines (black), a functional fit to the line profile (blue), and their spectro-astrometric signals (red) are plotted versus velocity.  The FWHM of the emission line is 19.51$\pm$0.27~km~s$^{-1}$. The blip near --40~km~s$^{-1}$ is the location of the low-J CO telluric lines. Since this region is excluded from the average for the low-J lines, the SNR of this part of the average spectrum is lower. The SNR of the continuum is 520 and the pixel to pixel standard deviation of the spectro-astrometric signal is 0.44~mas.  }
        \label{fig:3}
\end{figure*}

\begin{figure*}
        \includegraphics[width=0.9\textwidth]{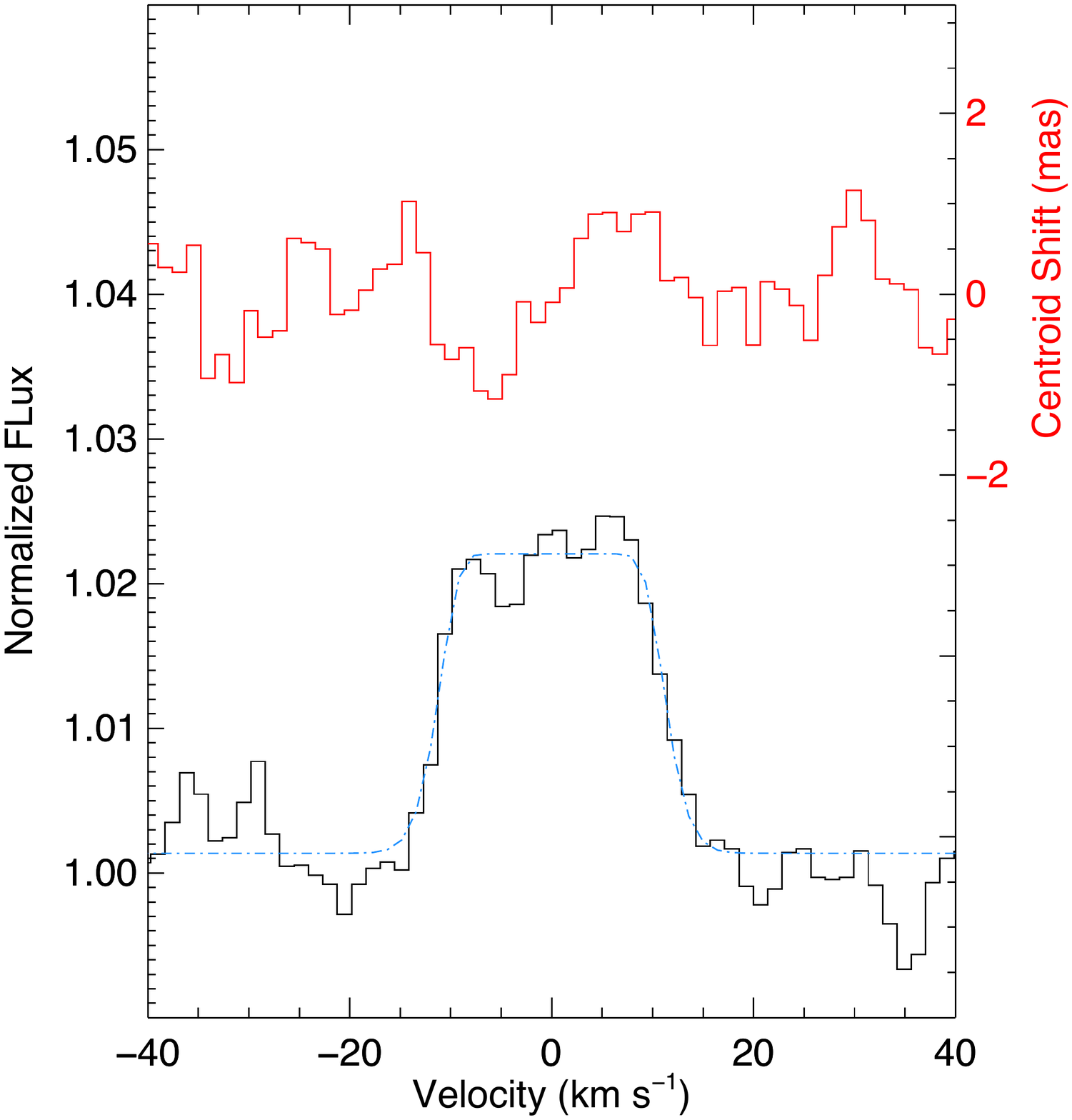}
        \caption{Profile of v=2--1 CO lines and spectro-astrometric signal. The average profile of the v=2--1 spectral lines (black), a functional fit to the line profile (blue), and their spectro-astrometric signals (red) are plotted versus velocity. The FWHM of the emission line is 22.64$\pm$0.66~km~s$^{-1}$. } 
        \label{fig4}
\end{figure*}

\begin{figure*}
        \includegraphics[width=0.9\textwidth]{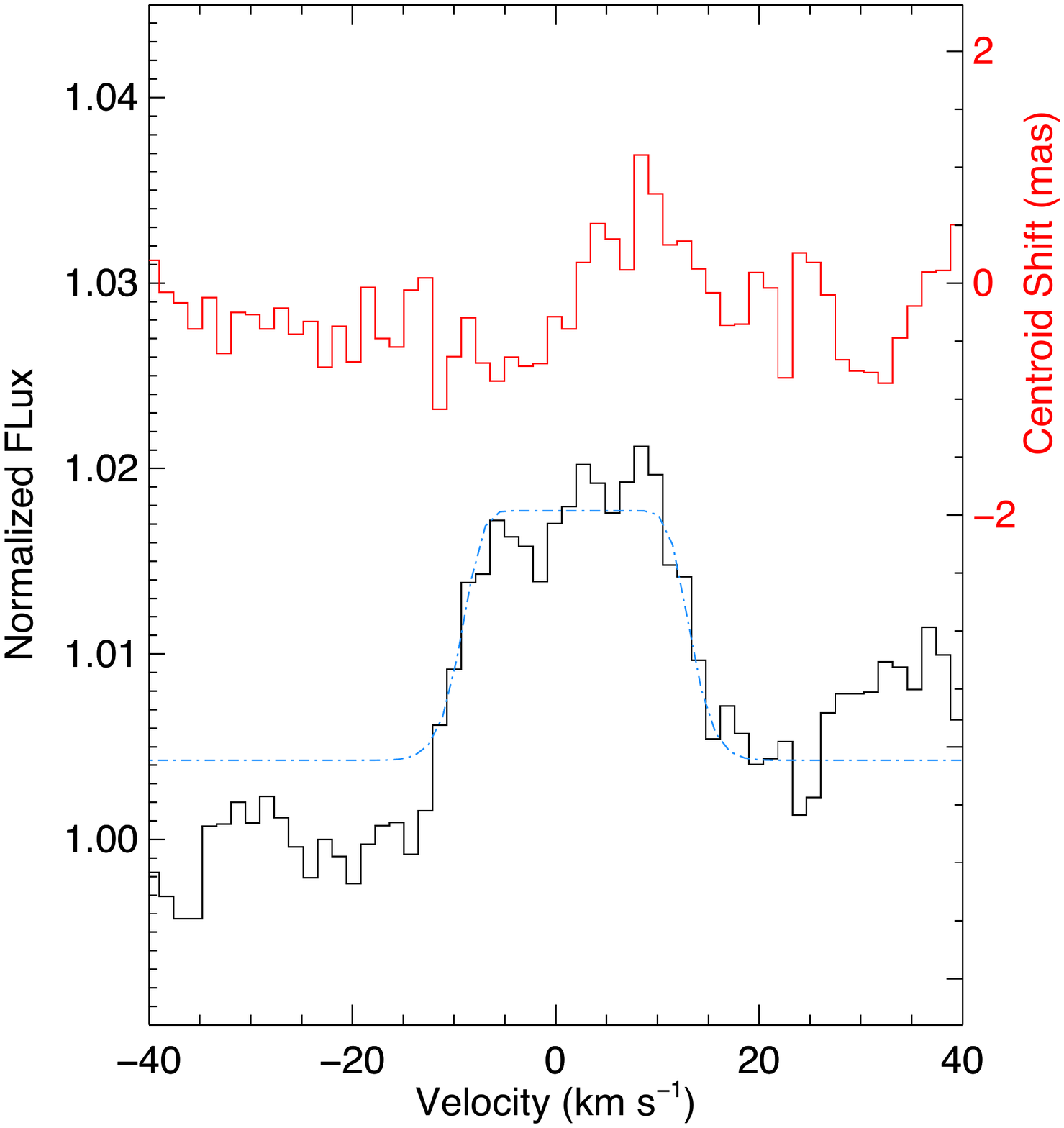}
        \caption{Profile of v=3-2 CO lines and spectro-astrometric signal. The average profile of the v=3--2 spectral lines (black), a functional fit to the line profile (blue), and their spectro-astrometric signals (red) are plotted versus velocity. The FWHM of the emission line is 22.17$\pm$1.07~km~s$^{-1}$.} 
        \label{fig5}
\end{figure*}

\begin{figure*}
        \includegraphics[width=0.9\textwidth]{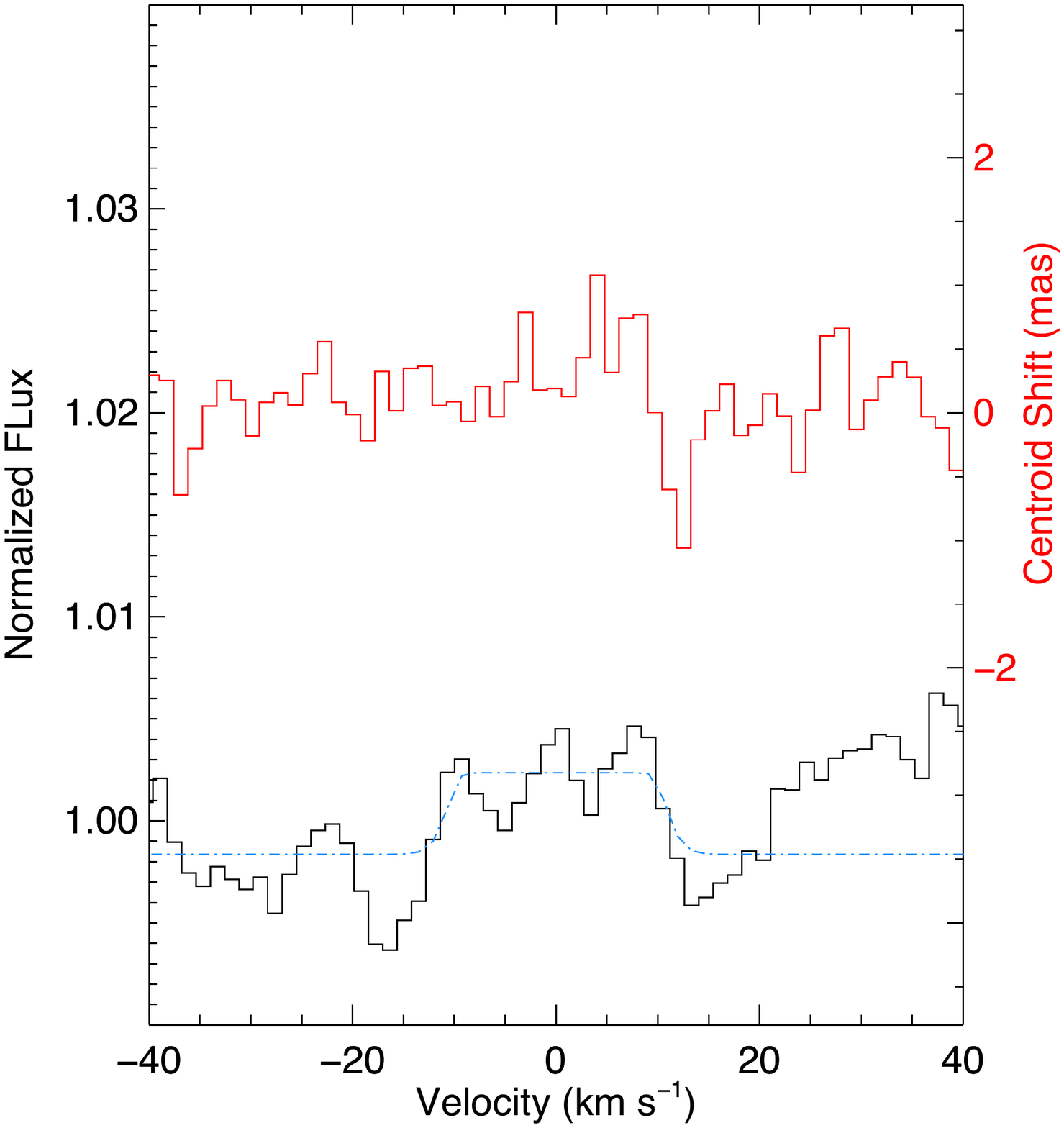}
        \caption{Profile of v=4--3 CO lines and spectro-astrometric signal. The average profile of the v=4--3 spectral lines (black), a functional fit to the line profile (blue), and their spectro-astrometric signals (red) are plotted versus velocity. The v=4--3 lines are not detected.} 
        \label{fig6}
\end{figure*}

\begin{figure*}
        \includegraphics[width=0.9\textwidth]{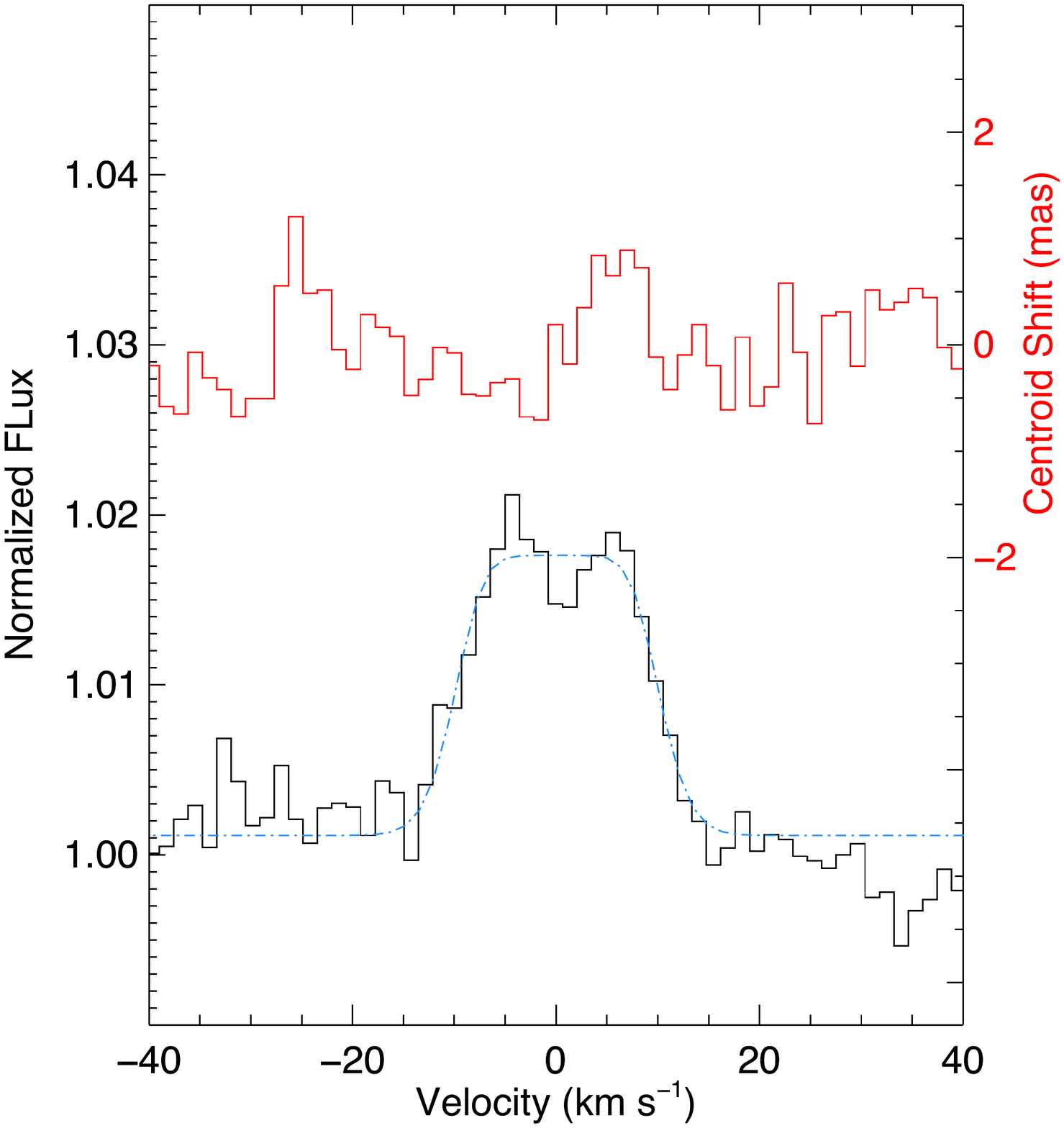}
        \caption{Profile of $^{13}$CO lines and spectro-astrometric signal. The average profile of the v=1--0 spectral lines (black), a functional fit to the line profile (blue), and their spectro-astrometric signals (red) are plotted versus velocity. The FWHM of the emission line is 20.15$\pm$0.91~km~s$^{-1}$.} 
        \label{fig7}
\end{figure*}

\begin{figure*}
        \includegraphics[width=0.95\textwidth]{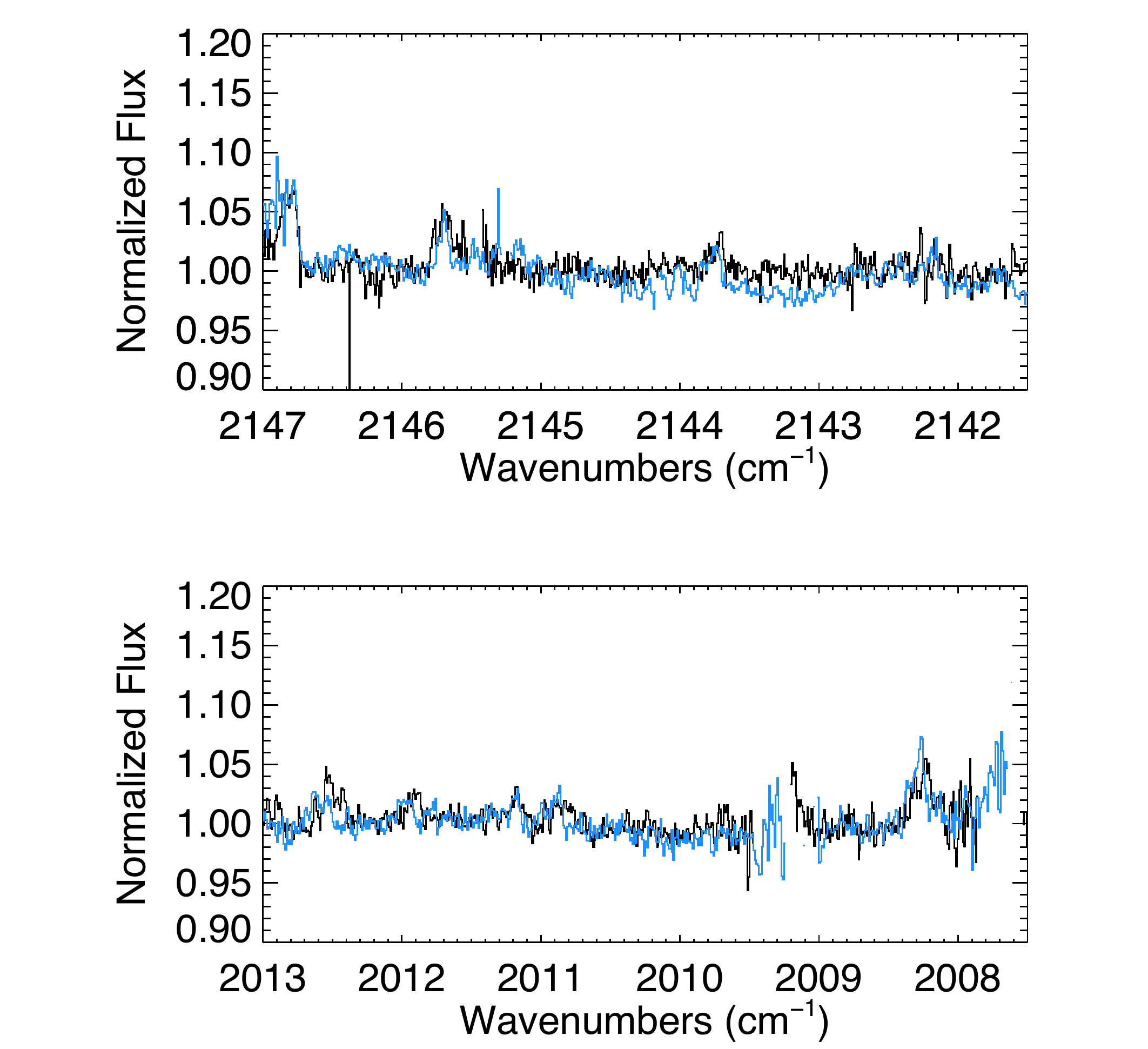}
        \caption{Comparison of spectrum acquired with CRIRES. The spectrum acquired with iSHELL (black) is plotted with the spectrum acquired with CRIRES (blue). These regions were selected to minimize systematic features from miscanceled telluric lines. The per pixel SNRs of the continua near 2145cm$^{-1}$ and 2011.5cm$^{-1}$ are 120-130 in the iSHELL spectrum and the CRIRES spectrum. The iSHELL spectrum was the result of 30 minutes of integration. The integration times of the CRIRES spectra were 20 minutes and 16 minutes respectively.}
        \label{fig:9}
\end{figure*}

\begin{figure*}
        \includegraphics[width=0.85\textwidth]{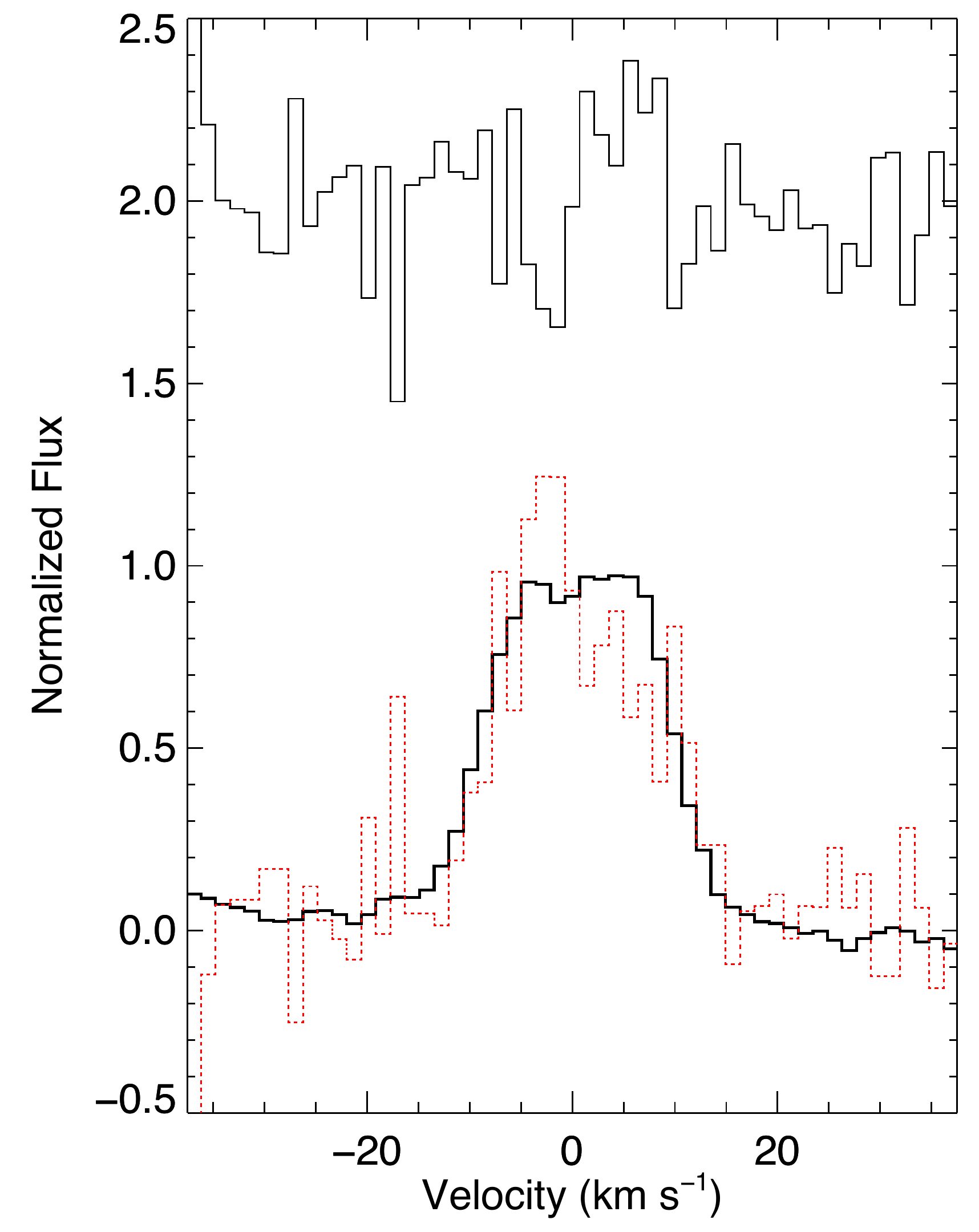}
        \caption{Comparison of the v=1-0 CO line profile constructed from data acquired with iSHELL and CRIRES. 
        The average profile of the v=1-0 lines observed with iSHELL (black) is plotted over the average profile of the v=1-0 lines observed with CRIRES (red). The line profiles have been scaled to a common equivalent width. The difference between the profiles is plotted above. The signal to noise ratio of the profile improved by a factor of 10 due to the large increase in the number of lines included in the average. The iSHELL spectrum was the result of 30 minutes of integration. The integration time of the CRIRES spectrum was 46 minutes. While there is some indication of an asymmetry in the profile constructed from lines observed with CRIRES, the signal to noise ratio is too low to conclude anything definitive. }
        \label{fig:8}
\end{figure*}

\begin{figure*}
        \includegraphics[width=0.95\textwidth]{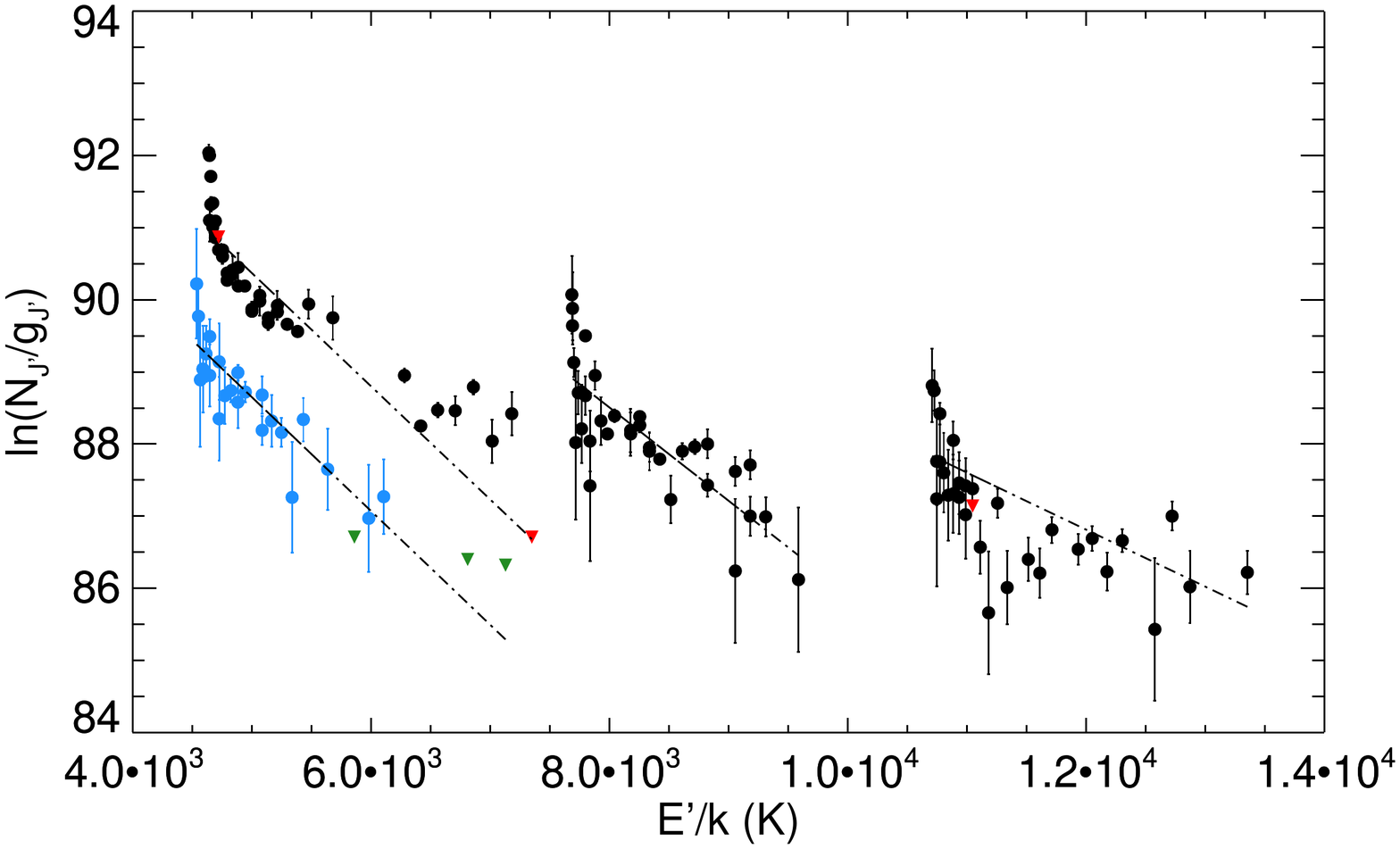}
        \caption{Excitation diagram of CO emission from HD~179218. E$^{\prime}$/k is plotted versus ln(N$_J^{\prime}$/g$_J^{\prime}$. The detected $^{12}$C$^{16}$O lines are plotted 
        with black points and the upper limits are plotted with red downward pointing triangles. The detected $^{13}$C$^{16}$O lines are plotted with blue points and the upper limits are plotted with green downward pointing triangles. There were no individual lines from the v=4--3 vibrational band detected. The rotational temperatures of the $^{12}$C$^{16}$O v=1--0, v=2--1, v=3--2, and $^{13}$C$^{16}$O v=1--0 lines are 640$\pm$10~K, 770$\pm$30~K, 1270$\pm$120~K, and 630$\pm$90~K respectively. The vibrational temperature is 1850$\pm$90~K.}
        \label{fig:10}
\end{figure*}

\end{document}